\documentclass[twoside]{dis09}
\usepackage[latin1]{inputenc}
\usepackage[dvips]{graphicx,epsfig,color}
\usepackage{wrapfig,rotating}
\usepackage{amssymb,amsmath,array}

\begin{document}
\title{$K_s^0$ Hadronization Following DIS at CLAS}

\author{K. Hicks and A. Daniel
%
\thanks{Supported in part by NSF grant PHYS-0653454}
%
\vspace{.3cm}\\
Department of Physics and Astronomy \\
Ohio University, Athens, Ohio, 45701, USA
}

\maketitle

\begin{abstract}
The hadronization of $K^0$ particles was measured in semi-inclusive 
deep inelastic scattering (SIDIS) kinematics for several nuclear 
targets using the CLAS detector.  Multiplicity ratios and 
$\Delta p_T^2$ values were extracted from the data.  These results 
may be compared with similar values for $\pi^+$ hadronization 
from CLAS (at the same kinematics) and $K^+$ hadronization from 
HERMES (at higher energy transfer). The physics goal of these 
measurements is to understand the space-time evolution as the 
struck quark becomes a full-blown hadron as it propagates through 
nuclear matter.
\end{abstract}

\section{Introduction}

The CLAS experiment called {\em eg2} was run with a variety of nuclear 
targets using a 5.5 GeV electron beam from the continuous electron beam 
accelerator facility (CEBAF) at Jefferson Lab \cite{eg2}.
The goal of this experiment is to measure observables related to 
the propagation of a quark struck by the virtual photon from 
deep inelastic scattering (DIS) through cold nuclear matter. 
These results will be contrasted with quark propagation through hot 
QCD matter, sometimes called the quark-gluon plasma (QGP), formed in 
relativistic heavy ion collisions (RHIC). Quarks from RHIC are 
tagged by back-to-back ``jets" of high energy hadrons formed by 
hard quark-quark collisions.  In contrast with RHIC results, where 
the quark propagation is severely damped by passage through the QGP,
the DIS data from HERMES \cite{hermes} suggest that the struck quark 
propagates more easily (compared with fully formed mesons) 
through cold nuclear matter.  

The measured quantities of the SIDIS measurements are the squared 
four-momentum transfer ($Q^2$) and energy transfer ($\nu$) from 
the scattered electron and the energy fraction ($z = E_h/\nu$) 
and transverse momentum ($p_T$) of the leading hadron.  By comparing 
the $z$ and $p_T$ distributions of pions and kaons from a deuterium 
target with that from various nuclear targets, the quark propagation 
and formation into a hadron can be inferred.  Theoretical predictions 
in a given model provide guidance to interpret the experimental results.
In this way, the space-time characteristics of QCD (in the process 
of forming a hadron from the struck quark) becomes accessible
\cite{brooks}. 

Ratios of the number of hadrons detected, normalized to the number 
of DIS events in a given kinematic bin of $Q^2$ and $\nu$, for a 
nuclear target divided by the same quantities for a deuterium target 
are one quantitative measure of the effects of propagation through 
cold nuclear matter.  The measured multiplicity ratio is given by:
$$ R = \frac{ (N_h(\nu,Q^2,z)/N_e(\nu,Q^2))_A }
            { (N_h(\nu,Q^2,z)/N_e(\nu,Q^2))_D } $$
where $N_h$ is the number of hadrons of type $h$ (here, $h$ is a pion 
or a kaon) with energy fraction $z=E_h/\nu$ measured for semi-inclusive 
final states (SIDIS), $N_e$ is the number of electrons in DIS, and 
the numerator is for a given nuclear target $A$ while the denominator 
is for deuterium $D$.  
Because $R$ is a super-ratio (a ratio of ratios), 
many systematic effects cancel.  For example, the detector 
efficiency to measure an electron at a given kinematics ($\nu$ and $Q^2$) 
will be the same (to first order) for targets $A$ and $D$.  Similarly, 
the detection efficiency of a pion with energy fraction $z$ in a 
given kinematic bin will be the nearly same for targets $A$ and $D$.
Small effects due to differences in the geometry of the two targets 
can be corrected using Monte Carlo simulations.  The interpretation 
of the multiplicity ratio in terms of the physics of the formation 
time of a given hadron $h$ can be obtained using theoretical models 
such as the BUU model \cite{Mosel}.

Another observable, the difference of the mean-square of the transverse 
momentum of a hadron, $\langle p_T^2 \rangle$, measured from the axis 
of the virtual photon momentum $\vec{q}$, 
for a nuclear target minus that for a deuterium target is given by 
$$ \Delta p_T^2 = \langle p_T^2 \rangle_A 
                - \langle p_T^2 \rangle_D  \ . $$
This quantity, according to theoretical models, is sensitive 
to gluonic radiation by the quark before it forms into a hadron
\cite{BDMPS}. 
This can be plotted as a function of $\nu$ for various nuclei, $A$, 
and can be compared with theoretical predictions \cite{BDMPS} of the
quark gluonic energy loss,
\begin{equation}
\label{BDMPSdEdx}
\left( \frac{dE}{dx} \right)_{medium} = \frac{3}{4}\alpha_s \Delta p_T^2
\end{equation}
where $\alpha_s$ is the strong coupling constant.
The mechanism of energy loss is primarily medium-stimulated gluon
radiation, although collisional losses are also treated. Although
Ref. \cite{BDMPS} considers `cold' nuclear matter they do not 
specifically address data from nuclei, as was done by Ref. \cite{Wang2}.

\section{Preliminary Results}

\begin{wrapfigure}{r}{0.6\columnwidth}
\centerline{\includegraphics[width=0.55\columnwidth]{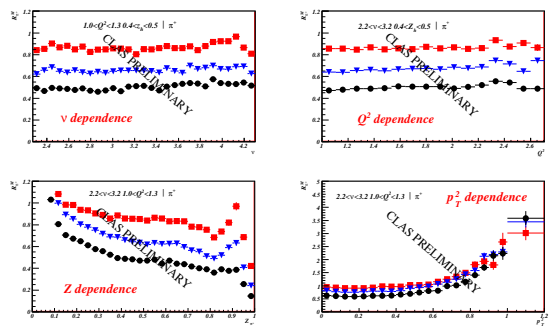}}
\caption{Preliminary results for kinematic dependence of the 
super-ratio $R$ for $\pi^+$ hadronization detected in CLAS
(updated analysis from Ref. \cite{brooks}).}
\label{fig:piplus}
\end{wrapfigure}

By comparing these two quantitative measures with theoretical 
calculations over a variety of kinematics, models of hadronization 
can be tested.  The HERMES experiment has already published the 
nuclear attenuation ratios for several types of hadrons at DIS 
kinematics in the range $8<\nu<23$ GeV \cite{hermes}.  
The CLAS data are for a lower energy range, with $2<\nu<5$ GeV, 
corresponding to shorter hadron formation times.  
At HERMES kinematics, calculations suggest \cite{hermes} that 
the quark propagates through the full nucleus, followed by hadron 
formation outside of the nuclear radius.  At CLAS, we expect to 
see hadron formation inside the nuclear radius, at least for 
heavier nuclear targets (larger $A$).

\begin{wrapfigure}{r}{0.5\columnwidth}
\centerline{\includegraphics[width=0.4\columnwidth]{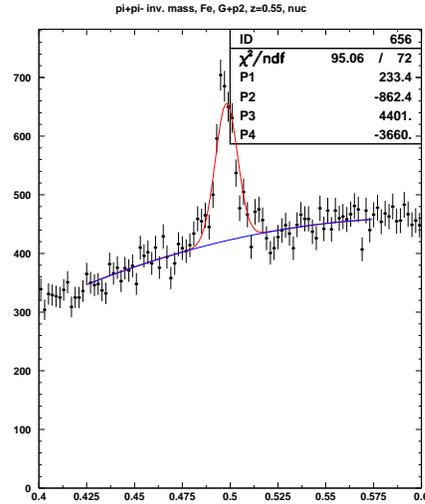}}
\caption{Invariant mass of a $\pi^+\pi^-$ pair 
detected in CLAS, showing the $K_s^0$ peak.}
\label{fig:k0peak}
\end{wrapfigure}

Preliminary results for $\pi^+$ produced at CLAS \cite{brooks}, 
shown in Fig.~\ref{fig:piplus}, indicate little dependence 
of the multiplicity ratio $R$ on the DIS variables $\nu$ and $Q^2$ 
for a given nuclear target, whereas $R$ does depend on $z$ and $p_T^2$. 
Since the attenuation of $R$ does depend on the nucleus
at these kinematics, the results of Fig.~\ref{fig:piplus}  
suggests that the average hadron formation 
length is shorter than the radius of Pb.  Nuclear 
multiplicity ratios cannot be explained by theoretical models unless 
some finite propagation of a "pre-hadron" (with a smaller cross 
section than fully-formed hadrons) is assumed \cite{Mosel}.  

Here we present for the first time results for $K_s^0$ hadronization 
measured in SIDIS kinematics at CLAS.  The $K^0$ was detected by 
reconstructing the invariant mass of a $\pi^+\pi^-$ pair from the 
$K_s^0$ decay branch.  
A sample mass spectrum showing the $K_s^0$ peak is shown in 
Fig.~\ref{fig:k0peak} for the Fe target in the energy fraction bin 
spanning $z=0.5$-0.6.  
The number of $K_s^0$ particles was obtained by a gaussian fit 
over a smooth polynomial background. 
Similar fits were done for the liquid deuterium LD$_2$ target. 

\begin{wrapfigure}{r}{0.5\columnwidth}
\centerline{\includegraphics[width=0.45\columnwidth]{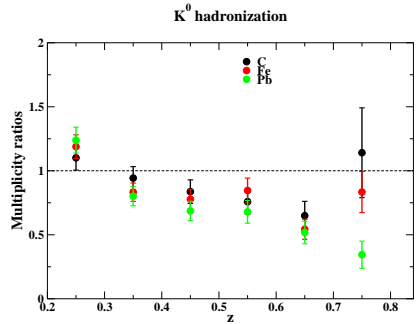}}
\caption{ Preliminary multiplicity ratios for the targets shown for 
$K_s^0$ hadronization summed over the kinematics of eg2.
}
\label{fig:mrat}
\end{wrapfigure}

To get the multiplicity ratio, the number of $K_s^0$ events for SIDIS 
events were normalized by the number of DIS events with 
of $Q^2 > 1.0$ GeV$^2$ and $W>2$ GeV.  Unlike the CLAS 
$\pi^+$ hadronization data \cite{brooks}, the statistics for $K_s^0$ 
were not sufficient to obtain multiplicity ratios for different bins 
in $Q^2$ and $\nu$.  Ratios of SIDIS/DIS were obtained for several 
targets (LD$_2$,C,Fe,Pb) and the multiplicity ratio $R$ was 
calculated for a given nuclear target normalized by the deuterium 
ratio.  Small corrections for the detector efficiency from the 
difference in geometry of the nuclear and LD$_2$ targets were 
done with Monte Carlo simulations.  Preliminary results are shown in 
Fig.~\ref{fig:mrat} as a function of the energy fraction $z$ of 
the $K_s^0$.  

Although the data in Fig.~\ref{fig:mrat} are still preliminary, 
there is an indication that the multiplicity ratios are less attenuated 
(closer to unity) than for $\pi^+$ hadronization at the same 
kinematics.   This in turn suggests that there is a dependence 
on the mass of the antiquark picked up by in the hadronization 
process of a meson, at least for $z>0.5$ where the meson is 
likely to include the struck quark (and an antiquark from the 
breaking of the color flux tube).

At the highest $z$-bin in Fig.~\ref{fig:mrat}, $z=0.7$-0.8, the 
statistical uncertainties (shown by the error bars) are large, 
but there is a hint that the multiplicity ratio for carbon and iron 
targets diverges from the downward trend seen at lower $z$.  A 
similar effect was seen in the higher-statistics $\pi^+$ data 
above $z=0.8$ shown in Fig.~\ref{fig:piplus}.  The reason for 
this behavior in $z$ is not known at the present time, however 
at the highest $z$ the hadronization data are known to be dominated 
by quasifree resonance production at the CLAS kinematics. For 
the $K_s^0$ data (Fig.~\ref{fig:mrat}), the multiplicity ratio 
for $z>0.8$ are not shown because quasifree resonance production 
is not the desired reaction mechanism in the calculation of $R$.

\begin{wrapfigure}{r}{0.5\columnwidth}
\centerline{\includegraphics[width=0.45\columnwidth]{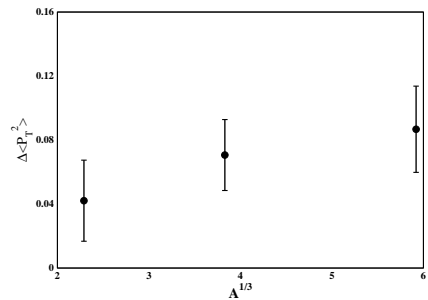}}
\caption{ Preliminary results of $\Delta p_t^2$ for $K_s^0$ mesons 
produced by nuclear targets $A$, as a function of $A^{1/3}$.
}
\label{fig:pt2}
\end{wrapfigure}

Preliminary results of the average transverse momentum broadening, 
$\Delta p_T^2$, are shown in Fig.~\ref{fig:pt2} for $K_s^0$ hadronization. 
Again, the statistical uncertainties are large, even after summing 
over the kinematics of the CLAS-eg2 experiment, but it appears 
that $\Delta p_T^2$ is broadened (greater than zero) as expected. 
The data in Fig.~\ref{fig:pt2} show a hint of increased $\Delta p_T^2$ 
for larger nuclei $A$, however the value for the three targets 
is consistent within error bars.  For the higher-statistics 
$\pi^+$ data shown in Fig.~\ref{fig:piplus}, a clear dependence 
of $\Delta p_T^2$ is seen on the nuclear target. 

As explained above, the value of $\Delta p_T^2$ is sensitive (in 
theoretical calculations) to gluonic radiation as the struck quark 
propagates through the nucleus, before forming a hadron.  Comparison 
of the value of $\Delta p_T^2$ for different mesons (such as pions 
and kaons) are of interest to theorists.  It is possible that the 
propagation time of the struck quark (before the breaking of 
the color flux tube) could be different depending on the mass 
of the antiquark in a given meson.

Clearly, higher statistics for $K_s^0$ hadronization are necessary 
before any conclusions can be drawn from theoretical interpretations 
of the $\Delta p_T^2$ data from CLAS-eg2.  Partly for this reason, 
a proposal was approved to study hadronization at the planned 
upgrade to 12 GeV electron beam energy at Jefferson Lab 
\cite{upgrade}. The CLAS detector will also be upgraded to handle 
the higher-momentum particles from the 12 GeV beams, with 
particular emphasis on DIS kinematics with higher luminosity 
(up to $10^{35}$ cm$^{-2}$s$^{-1}$).  With this upgrade, a 
wider range of DIS kinematics will be accessible by CLAS12. 
At the higher beam energies, kaon production is known to 
have higher rates of production, allowing greater statistical 
accuracy. 


\section{ Summary }

In conclusion, the CLAS data provide a new kinematic window for 
hadronization at the expected length scale near to the radius of 
heavy nuclei.  Statistical uncertainties for $K_s^0$ hadronization 
in DIS kinematics at CLAS-eg2 are still large, even though the 
CLAS-eg2 luminosity is larger than for the HERMES data by a factor 
of $\sim 100$, in part because of the low production rates for 
$K^0$ mesons at lower average $\nu$. Nonetheless, the preliminary 
data shown here are the only available data for $K_s^0$ hadronization 
and hence are a first step toward extending the HERMES results.
Both $\pi^+$ and $K_s^0$ multiplicity ratios and $\Delta p_T^2$ 
are expected to be finalized in the near future, as more data 
from the eg2 run becomes available.

\section{Acknowledgments}

Discussions with Will Brooks, the eg2 Run Group and the CLAS Collaboration 
are gratefully acknowledged.


\begin{footnotesize}

\end{footnotesize}


\end{document}